%% file: IET-Conf-Paper-sample.tex
\documentclass{IET-Conf-Paper}

{}
{}
{}

\usepackage{amsmath,amssymb,amsfonts}
\usepackage{algorithmic}
\usepackage{graphicx}
\usepackage{textcomp}
\usepackage{xcolor}
\usepackage[printonlyused,nolist]{acronym}
\usepackage{tikz}
\usetikzlibrary{shapes.geometric, arrows}
\usepackage{hyperref} %for linking the cross references\usepackage{graphics} %for dealing with figures.
\usepackage{algorithmic} %for describing algorithms
\usepackage{authblk}
\usepackage{multicol}
\usepackage{blindtext}
\usepackage{epstopdf}
\usepackage{float}
\usepackage{listings}
\begin{document}

\title{\vspace{-2em}Graph-based Impact Analysis of Cyber-Attacks on Behind-the-Meter Infrastructure
}

\author{Immanuel Hacker\ad{1,2}\corr, Ömer Sen\ad{1,2}, Florian Klein-Helmkamp \ad{2}, Andreas Ulbig\ad{1,2} }

\address{\add{1}{Digital Energy Fraunhofer FIT, Aachen, Germany}
\add{2}{IAEW at RWTH Aachen University, Aachen, Germany}
\email{immanuel.hacker@fit.fraunhofer.de}}


\begin{abstract}
Behind-the-Meter assets are getting more interconnected to realise new applications like flexible tariffs. Cyber-attacks on the resulting control infrastructure may impact a large number of devices, which can result in severe impact on the power system.
To analyse the possible impact of such attacks we developed a graph model of the cyber-physical energy system, representing interdependencies between the control infrastructure and the power system. This model is than used for an impact analysis of cyber-attacks with different attack vectors.
\end{abstract}

\maketitle

\vspace{-2em}
\begin{center}
    \textit{This paper is a preprint of a paper submitted to 14th Mediterranean Conference on Power Generation Transmission, Distribution and Energy Conversion (MEDPOWER 2024) and is subject to Institution of Engineering and Technology Copyright. If accepted, the copy of record will be available at IET Digital Library}
\end{center}
\vspace{-1em}

\input{acronyms}
\input{tixz_style}

\input{lst_style}

\section{Introduction}
The energy transition has led to two developments that increase the significance of \ac{BTM} infrastructure. On the one hand, electrification of the heating and mobility sectors, and on the other, decentralised energy production through \ac{PV} systems. Additionally, storage systems are increasingly integrated to enable economically optimal combined utilisation of production and consumption. These \ac{BTM} installations are interconnected through \ac{ICT} infrastructure often via the internet, due to the requirements from various stakeholders.
The applications that necessitate this connectivity are highly diverse.
The simplest example is the customer's need to remotely control activities such as charging their own \ac{EV} or heating their house.
However, more complex applications such as dynamic electricity tariffs, which control \ac{BTM} devices based on market prices, are already a reality.
From the perspective of network operators, due to the new challenges for the electricity grid, it is crucial that \ac{BTM} installations can be controlled in a  grid-serving manner to ensure the safe operation of the electricity grid at all times.
These applications are either implemented directly via the customer's internet connection using \ac{HEMS} or through a dedicated advanced metering infrastructure.
The interconnection of the installations is imperative to implement these applications.
However, the networking of many small assets via a common \ac{ICT} infrastructure or a common software stack carries the risk that in the event of a cyber-attack, a large number of assets may be simultaneously affected, which could cumulatively have a significant impact on the power system.
Therefore, we propose a method to quantitatively assess this impact for various attack scenarios.

\subsection{Related Work}
\label{subsec:furtherwork}
The goal of this paper is to use approaches from the \ac{SW} to build the graph model, therefore the related work section focuses on  
\ac{SW} technologies are being explored in many areas for data modelling and describing semantic relationships between entities and attributes. Much research has also focused on how this technology can be connected to smart grid research areas.
%\todo{CGMS abgernzung einbauen für Produktiv Systeme verschiedenenn usecases im berreich Forschung aandere anforderung an Daten oft deutlich geruuinger manchmal anders -> perspektifich interoperabilität}
Several approaches exist to build ontologies for power systems using \ac{RDF}.
For instance, \cite{devanand_ontopowsys_2020} focuses on multi-domain energy systems, while \cite{chun_knowledge_2018} concentrates on micro-grid communities and on the service side of the system.
Both use \ac{SW} technologies, but not with the purpose of bridging different domains and enable cyber resilience research.
The proposed models do not consider a holistic framework for describing smart grid use cases with all associated layers in both cases. In particular, they lack the connection to the \ac{SGAM}, which brings the concept in line with a widely used way to describe use cases, making the models much more accessible for researchers.
\cite{sandbergdeliverable} uses the \ac{SGAM} as basis for an ontology designed for threat modelling.
Although this work involves modelling cross-domain inter-dependencies of cyber-physical systems, it does not utilise SHACL, which allows validation through rules.
In contrast to the work presented, our paper proposes an approach that provides a holistic picture that presents a complete framework and shows how \ac{SW} technologies can be useful in all steps of definition, description, validation, and evaluation.
One of these additional aspects is a rule-based augmentation for transforming power systems in \ac{CPES}.
\cite{klaer_sg3_2020} presents an approach of graph-based modelling for \ac{CPES} and emphasises the need for an automated augmentation process. However, the work focuses only on a specific \ac{SCADA} use case and does not utilise \ac{SW} technologies. 

\input{methodology}
\input{workflow}

\input{casestudy}
\input{conclusion}

\section{Acknowledgment}
\noindent
\begin{minipage}{0.60\columnwidth}%
This work has received funding from the \ac{BMBF} under project funding reference 03SF0694A.
\end{minipage}
\hspace{0.02\columnwidth}
\begin{minipage}{0.35\columnwidth}%
    \includegraphics[width=\textwidth]{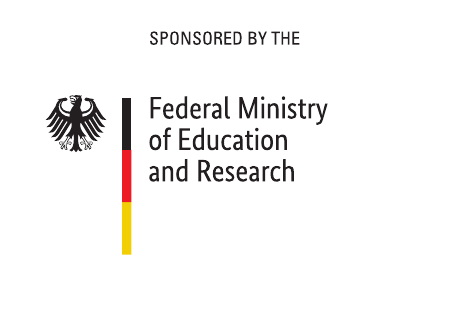}
\end{minipage}
\vspace{-1em}
\bibliographystyle{iet}
\bibliography{references.bib}

\end{document}

%% file: acronyms.tex
\begin{acronym}
	\acro{API}{Application Programming Interface}
	\acro{AMI}{advanced metering infrastructure}
	\acro{BMWK}{German Federal Ministry for Economic Affairs and Climate Action}
    \acro{BMBF}{Federal Ministry of Education and Research}
	\acro{BUM}{Bottom-Up Model}
 	\acro{BTM}{Behind-the-Meter}
	\acro{BSS}{Battery Storage System}
    \acro{CWA}{Closed World Assumption}
    \acro{CVE}{Common Vulnerabilities and Exposures}
    \acro{CPES}{Cyber Physical Energy System}
    \acro{DC}{Direct Current}
  	\acro{DER}{Decentralized Energy Resource}
	\acro{DSM}{Demand Side Management}
	\acro{DSO}{Distribution System Operator}
	\acro{DoS}{Denial of Service}
	\acro{DMZ}{Demilitarized Zone}
    \acro{EV}{Electric Vehicle}
	\acro{ET}{Energy Technology}
	\acro{EMS}{Energy Management System}
    \acro{ERROL}{Energy Resilience Research Ontology Language}
 	\acro{FDI}{False Data Injection}
    \acro{HEMS}{Home Energy Management System}
	\acro{RDF}{Resource Description Framework}
    \acro{RDFS}{RDF Schema}
	\acro{ICT}{Information and Communications Technology}
	\acro{IDS}{Intrusion Detection System}
	\acro{IT}{Information Technology}
 	\acro{IED}{Intelligent Electronic Device}
    \acro{KG}{Knowledge Graph}
 	\acro{MTU}{Master Terminal Unit}
	\acro{MitM}{Machine in the Middle}
    \acro{MV}{Medium Voltage}
	\acro{OT}{Operational Technology}
    \acro{OWA}{Open World Assumption}
    \acro{PCAP}{Packet Capture}
    \acro{PE}{Privilege Escalation}
	\acro{PV}{photovoltaic}
 \acro{PCC}{point of common coupling}
	\acro{RTU}{Remote Terminal Unit}
 	\acro{RCE}{Remote Code Execution}
    \acro{LCD}{Lowest Common Denominator}
	\acro{SCADA}{Supervisory Control and Data Acquisition}
	\acro{SMGW}{Smart Meter Gateway}
	\acro{SAM}{Simulated Attacker Model}
	\acro{SHACL}{Shape Constraint Langugage}
	\acro{SPARQL}{SPARQL Protocol and RDF Query Language}
	\acro{SUID}{set-user-ID}
	\acro{SGAM}{Smart Grid Architecture Model}
 	\acro{SG}{Smart Grid}
    \acro{SW}{Semantic Web}
    \acro{TSO}{Transmission System Operator}
    \acro{URI}{Uniform Resource Identifier}
	\acro{vRTU}{virtual Remote Terminal Unit}
	\acro{VPN}{Virtual Private Network}
	\acro{VPP}{Virtual Power Plant}
	\acro{VED}{Virtual Edge Device}

\end{acronym}

%% file: tixz_style.tex
\newlength{\myboxheigth}
\setlength{\myboxheigth}{0.5cm}
\newlength{\myboxwidth}
\setlength{\myboxwidth}{3cm}

\tikzstyle{data} = [rectangle, rounded corners, 
minimum width=\myboxwidth, 
minimum height=\myboxheigth,
text centered, 
draw=black, 
fill=red!30]

\tikzstyle{io} = [trapezium, 
trapezium stretches=true, % A later addition
trapezium left angle=70, 
trapezium right angle=110, 
minimum width=\myboxwidth, 
minimum height=\myboxheigth, text centered, 
draw=black, fill=blue!30]

\tikzstyle{process} = [rectangle, 
minimum width=\myboxwidth, 
minimum height=\myboxheigth, 
text centered, 
text width=\myboxwidth, 
draw=black, 
fill=orange!30]

\tikzstyle{decision} = [diamond, 
minimum width=\myboxwidth, 
minimum height=\myboxheigth, 
text centered, 
draw=black, 
fill=green!30]
\tikzstyle{arrow} = [thick,->,>=stealth]

%% file: lst_style.tex
\lstdefinelanguage{turtle}{
    basicstyle=\small,
    morekeywords={@prefix, a},
    morestring=[b][\color{blue}]",
    morecomment=[s][\color{gray}]{<}{>}
}

%% file: methodology.tex
\section{Modelling Approach}

The goal of the graph model is to not only describe the power system but also the \ac{ICT} infrastructure, and the functional design of the cyber-physical system.
The reason is that the resulting model should be a holistic approach for all aspects of the smart grid and should be expandable to future applications.
Furthermore, a goal of this model is to provide a basis for exchange scenarios in the research community.

\subsection{Smart Grid Architecture Model}
\label{subsec:sgam}

\begin{figure}[H]
	\centering
	\includegraphics[width=\linewidth]{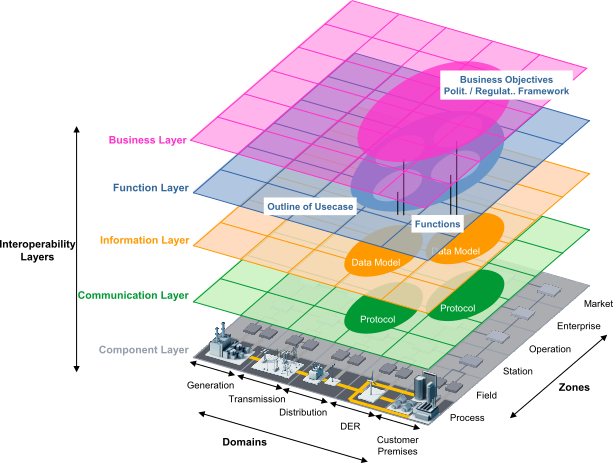}
	\caption{
		Visualisation of the Smart Grid Architecture Model showing the three-dimensional structure of the model\protect\cite{sgam2012}.
	}
	\label{fig:sgam}
\end{figure}

We chose to start off with an existing model, which offers a categorisation we can build up on and therefore ensure that an enhancement of the model happens in a consistent and interoperable way.
The \ac{SGAM} is categorised along three dimensions(Fig.: \ref{fig:sgam}) the \textit{Domains}, \textit{Zones} and, the \textit{Interoperability Layer}, the individual objects in our model are mapped to these dimensions\cite{sgam2012}.
The intent of \ac{SGAM} is a holistic approach to model use cases in smart grids.
Therefore, the original use case differs from the one in this paper as we want to model a system with instantiated models, not just a generic use case.
An advantage of this approach is that use cases, developed with \ac{SGAM}, can easily be transferred to models for analysis.
The main advantage of using \ac{SGAM} is its division into various interoperability layers.
This allows, for example, the examination of information flows independently of the protocols used, enabling a more universally applicable investigation. Similarly, the functions that different actors in an application have are considered independently of the hardware.
This stringent structuring allows for the impact analysis of different attack-vectors on various subareas of the cyber-physical energy system, such as a specific protocol or an attack on a particular functional actor.

\subsection{Closed World Assumption}
\label{subsec:cwa}
The \ac{OWA} and \ac{CWA} are two contrasting principles on how to model data.
Hereby the \ac{OWA} allows unknown or unrepresented data, so the absence of information is not considered evidence of absence and, therefore, not as invalid.
The \ac{OWA} is often used in \ac{SW} applications.
Still, because we need to verify that all the needed information for the later applications exists in the appropriate form, we follow the \ac{CWA} considering that a statement is true only if it is explicitly stated and all unknown information is considered false.
\ac{SHACL} allows us to follow the \ac{CWA}.

\section{Semantic Web Technologies}
\label{sec:technologies}

In the following, we will present \ac{SW} based technology, used in this paper.
\ac{SW} references the idea of a version of the internet where the information is linked and machine-readable so that the knowledge can be accessed and used by software tools.
To realise this vision, different technologies have been developed over the years, where the main challenges they tried to solve have been interoperability between different data sources and domains of knowledge.

\subsection{Resource Description Framework}
\label{subsec:rdf}

The \ac{RDF} is the main standard in the field of \ac{SW} to represent the information and the relationship between different objects.
Every information is represented by a triple consisting of a \textit{Subject}, \textit{Predicate} and, \textit{Object}.
An example would be \textit{"John hasAge 25"}: the resource John has the attribute age which has the value 25.
Instead of a value, the object can be another resource which creates a \ac{KG} representing the relationship between the different references.
Graphs have been proven to be a sophisticated way to model smart grids, in comparison to relational databases\cite{klaer_sg3_2020,jain_semantic_2023}.
The references used in \ac{RDF} are \ac{URI}s, so the linkage of data even outside the \ac{KG} is possible.
\ac{RDFS} are a mechanism in \ac{RDF} to define classes, properties and, basic constraints in \ac{KG}, therefore allowing us to build class hierarchies.
Because of the wide usage of \ac{RDF}, frameworks and tools for handling the data and automated reasoning are widely available.
As well as a high compatibility with different data formats for storage like \textit{XML} and \textit{TURTL}, listing \ref{lis:statgen} shows an extract from such a \textit{TURTL} file.

\subsection{SPARQL Protocol And RDF Query Language}
\label{subsec:sparql}

\ac{SPARQL} is designed to interact with \ac{RDF} graphs and supports simple and complex graph pattern matching which enables sophisticated querying capabilities to analyse relationships between objects.
In addition to information extraction, \ac{SPARQL} also allows graph modification by adding, deleting, and changing data. 

\subsection{Shapes Constraint Language}
\label{subsec:shacl}

An important challenge of the \ac{SW} is to ensure data quality and interoperability, to tackle this, constraints can be specified and validated with \ac{SHACL}.
Although \ac{RDFS} already provides capabilities for simple constraints,  SHACL provides a formal syntax and vocabulary for expressing these constraints and offers a more expressive and flexible validation mechanism compared to \ac{RDFS}.
Shapes define the expected patterns and constraints, which can be rules like required properties, data types and, value ranges.
An example where these are more advanced than \ac{RDFS} is checking for cardinality and value ranges which is essential for the \ac{KG} we want to build.
Because the constraints are defined separately from the data graph, different shapes can be applied for the validation of a data graph depending on the context of the investigation, for example, if a detailed model of the communication network is needed or just abstract communication channels.

\section{Ontology}
\label{sec:data_model}

The ontology section outlines the way we modelled the data for this paper, split into the domains: power system, communication network and operational technology.
To do so, we show components of the ontology based on \ac{SGAM}, entailing the interoperability layers which we use to organise the \ac{KG}.
Using this approach enables the independent design of the interoperability layers for rapid prototyping and error detection.

\subsection{Power System}
\label{subsec:powersystem}
The degree of detail that we envision is to model the component layer describing the power system with power flow calculations in a stationary time horizon.
Therefore we begin by modelling the power system in sufficient detail to enable these simulations, the data is mainly in the \textit{Component Layer}.
However, it is important to note that the model can be enhanced in the future to address additional needs.
We started off by modelling the power system based on the data model of \textit{PandaPower}\cite{pandapower_2018} as one of the most common open source frameworks used, so the compatibility to import models from other tools is quite high.
The original data model is a relational model, in contrast to the graph model we use, which closely resembles the character of the system and allows us to use graph algorithms.
Furthermore \ac{RDFS} enables us to build a class hierarchy with inheritance and \ac{SHACL} lets us define specific constraints for the classes.
This already exhibits several benefits over using the original model because, as with the \ac{SHACL} based validation, errors in the \ac{KG} can be found, which wouldn't be checked otherwise, ensuring data quality.
Listing \ref{lis:statgen} shows the \textit{Turtl} representation of a static generator; it defines the parent class in \ac{RDFS} and the properties, including the allowed value range, for example, the active power has to be a negative value because positive values are defined as power consumption.

\begin{lstlisting}[language=turtle, caption={Representation of the \ac{RDFS} and \ac{SHACL} shapes of a static genarotor in \textit{Turtl}.}, label={lis:statgen}]
errol:StaticGenerator
  rdf:type rdfs:Class ;
  rdf:type sh:NodeShape ;
  rdfs:label "Static generator" ;
  rdfs:subClassOf errol:GenerationConsumption ;
  sh:and (
      [
        sh:path errol:p_mw ;
        sh:maxInclusive "0"^^xsd:decimal ;
      ]
    ) ;
  sh:property [
      rdf:type sh:PropertyShape ;
      sh:path errol:q_mvar ;
      sh:datatype xsd:decimal ;
      sh:description "reactive power" ;
      sh:maxCount 1 ;
      sh:minCount 1 ;
      sh:name "q mvar" ;
    ] ;
  sh:property [
      rdf:type sh:PropertyShape ;
      sh:path errol:type ;
      sh:datatype xsd:string ;
      sh:description "sgen type" ;
      sh:maxCount 1 ;
      sh:name "type" ;
    ] ;
  sh:targetClass errol:StaticGenerator ;
\end{lstlisting}

\subsection{Control Infrastructure}
\label{subsec:control}

\begin{figure}[H]
	\centering
	\includegraphics[width=\linewidth]{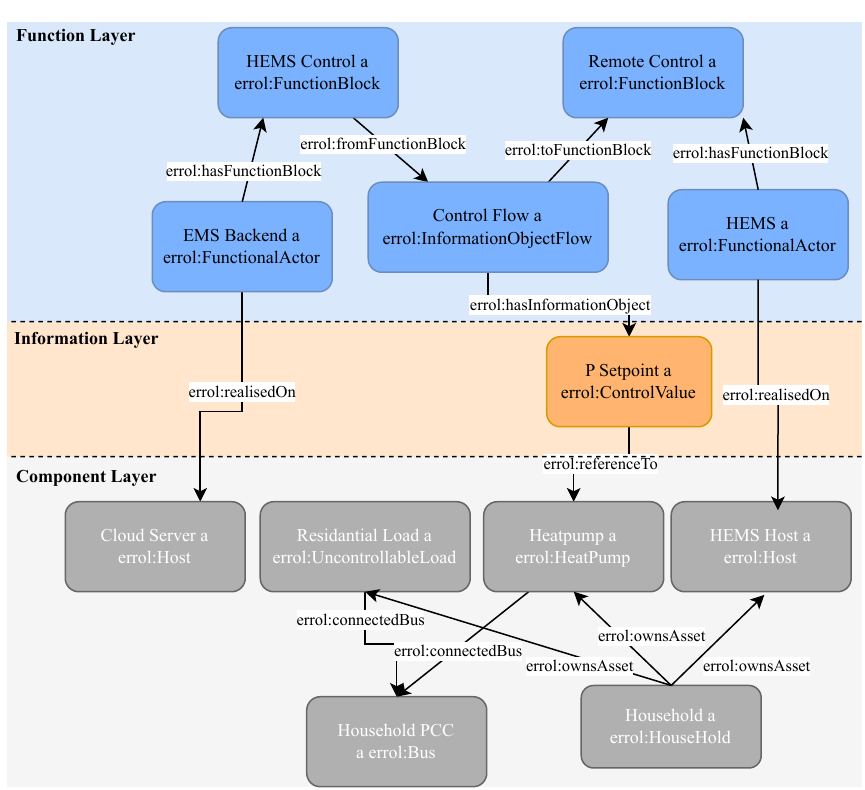}
	\caption{
		Simplified visualisation of the \ac{KG} used in this work, consisting of a \ac{HEMS} solution.
	}
	\label{fig:kg}
\end{figure}

The control infrastructure of the layers \textit{function}, \textit{Information}, and \textit{communication} but also of the physical \ac{ICT} infrastructure like hosts the systems are running on and the network components.
Because we manly focus on internet based solutions the internal infrastructure of the internet service provider is out of scope.
Figure \ref{fig:kg} shows the simplified \ac{KG} used in this paper for one household with a remote controllable heat pump.
\textit{errol}, the acronym for "Energy Resilience Research Ontology Library", is the namespace used for the newly developed ontology and indicates classes and predicates we defined.
\textit{errol:HouseHold} is a subclass of \textit{errol:AssetGroups}, which implements the idea of \textit{Zones} from \ac{SGAM} for an instantiated model, in this example an household is equivalent to \textit{station}.
Functional actors are elements which represent one functional entity in the system like a \ac{HEMS}.
Hereby the functions can be abstracted from the technical implementation, an functional actor may be implemented on a specific controller or may be implemented in a distributed manner.
\cite{neureiter_domain-specific_2017} describes a similar implementation to achieve the abstraction of the functional layer.
Functional blocks are the concrete implementations of individual functions of a functional actor.
A functional actor can hold multiple functional blocks simultaneously.
In a \ac{HEMS}, for example, one functional block might be self-consumption optimisation, while another functional block could be the remote controllability of a smart home.
In the example shown in figure \ref{fig:kg}, there is a functional actor for the cloud backend of the HEMS provider and another for the HEMS within the smart home. Each of these has a functional block that implements remote controllability.
The relationship between the functional blocks is realised through information object flows, which represent the flow of information and thereby the functional connections between the different actors.
\vspace*{-0.4cm}
These flows are directed, making the direction of the information flow clear.
As a result, control direction and monitoring must be represented in two separate information flows.
The transmission of measurement values has been omitted in this example, as it is not relevant for the intended impact analysis.
These elements are all part of the \textit{function layer}.
Each information object flow references the information objects that are being transmitted.
\textit{errol:ControlValue} is a subclass of \textit{errol:InformationObject}, specifically for control signals.
Information objects can stand alone or reference physical components, such as in this example, a heat pump whose power output can be controlled.
In this way, the information objects represent the relationship between the functional layer and the physical components.
Furthermore the hosts on which the functional actors are realised are added in the component layer, which allows for analysis of the dependency of applications on the physical \ac{ICT} infrastructure.

%% file: workflow.tex
\section{Workflow of Impact Analysis for Cyber-Attacks}

\begin{figure}[H]
\input{errol_flow}
    \centering
    \caption{The workflow starting from a \textit{PandaPower} model showing the different steps, including the augmentation step, impact analysis, and export.}
    \label{fig:errolflow}
\end{figure}
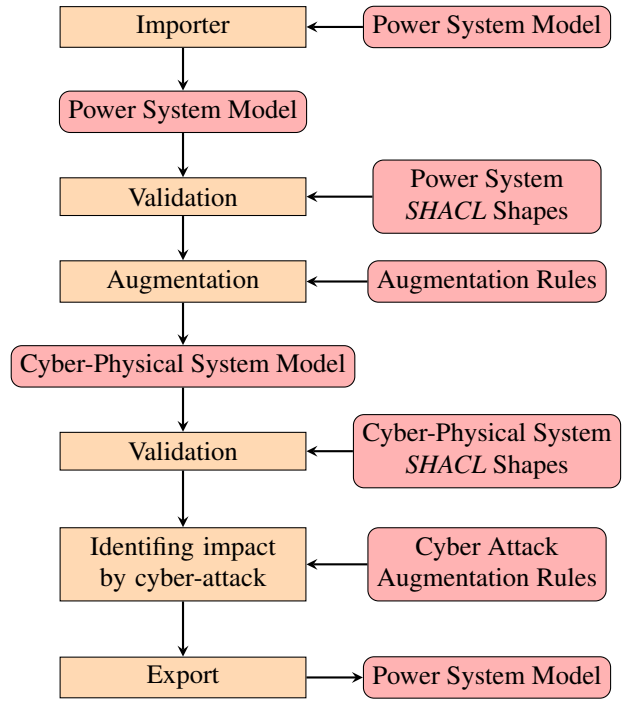

Figure \ref{fig:errolflow} shows the workflow of the impact analysis developed in this work.
The starting point for the investigations done in this work are \textit{PandaPower}\cite{pandapower_2018} models, which get imported into the presented graph format.
The power system model than gets validated against the \ac{SHACL} shapes for the power system.

Following this the control infrastructure is augmented automatically based on predefined augmentation rules.
This augmentation process contains the hardware of the control infrastructure as well the function blocks needed for the application and also the information flow between these function blocks.
Furthermore, information like the manufacturer and firmware version can be added to identify the scope of a cyber-attack which may only work on a specific firmware version.

The augmentation rules allow probabilistic distributions, for example equip 30 percent of electric vehicles with a specific \ac{HEMS} than connect these to a backend system depending on the manufacturer of the \ac{EV}.
\ac{SPARQL} is the foundation of the augmentation engine, which allows the definition of augmentation rules in an declarative way which which is a significantly more effective approach than solely imperative implementation the other parts are implemented in python.
The main advantage of this approach is the possibility to make probabilistic rules with probabilistic behaviour.
The augmentation rules are also realised as \ac{RDF} graph, bringing the same advantages of validation and re-usability.
Three kinds of rules are implemented \textit{add} rules allowing adding new object based on templates, \textit{change} rules allow to change or append a specific triple, and \textit{delete} rules.
A \ac{SPARQL} query is used to define which elements should be modified or to which element the new objects should be connected.
The probabilistic behaviour is defined by two probabilities the first defines if the rule is applied for specific element and the second defines which template should be used.
After the augmentation the result is a complete model of the cyber-physical energy system, which get validated by the corresponding \ac{SHACL} shapes.
The next step is the simulation of the cyber-attack, for this the type attack vector and the scope must be defined.
The cyber-attacks used for this paper are also implemented as \textit{change} rules in the augmentation engine.
An example may be a compromised backend of an \ac{EV} manufacturer allowing an attacker to arbitrarily control charging of connected \acp{EV}.
Based on the connections modelled in the graph model the change in the \textit{process zone} can be applied before exporting a power system model in the \textit{pandapower} format.
In the last step  power flow simulations are performed to examine the impact of different attack vectors on the power system.

%% file: errol_flow.tex
\begin{tikzpicture}

\node (pp) [data, align=center] {Power System Model};
\node (import) [process, left of=pp, xshift=-\myboxwidth]{Importer};
\node (pprdf) [data, align=center, below of=import, yshift=-0.25* \myboxheigth] {Power System Model};
\node (val1) [process, below of=pprdf, yshift=-0.25* \myboxheigth]{Validation};
\node (sh1) [data, align=center, right of=val1, xshift=\myboxwidth]{Power System\\\textit{SHACL} Shapes};
\node (aug) [process, below of=val1, yshift=-0.25* \myboxheigth]{Augmentation};
\node (cpsrdf) [data, align=center, below of=aug, yshift=-0.25* \myboxheigth] {Cyber-Physical System Model};
\node (rul) [data, align=center, right of=aug, xshift=\myboxwidth]{Augmentation Rules};
\node (val2) [process, below of=cpsrdf, yshift=-0.25 * \myboxheigth]{Validation};
\node (sh2) [data, align=center, right of=val2, xshift=\myboxwidth]{Cyber-Physical System \\\textit{SHACL} Shapes};
\node (impact) [process, below of=val2, yshift=-1 * \myboxheigth]{Identifing impact by cyber-attack};
\node (cyberrul) [data, align=center, right of=impact, xshift=\myboxwidth]{Cyber Attack\\Augmentation Rules};
\node (exp) [process, below of=impact, yshift=-1 * \myboxheigth]{Export};
\node (exppp) [data, align=center, right of=exp, xshift=\myboxwidth]{Power System Model};

\draw [arrow] (pp) -- (import);
\draw [arrow] (import) -- (pprdf);
\draw [arrow] (pprdf) -- (val1);
\draw [arrow] (val1) -- (aug);
\draw [arrow] (rul) -- (aug);
\draw [arrow] (sh1) -- (val1);
\draw [arrow] (aug) -- (cpsrdf);
\draw [arrow] (cpsrdf) -- (val2);
\draw [arrow] (sh2) -- (val2);
\draw [arrow] (val2) -- (impact);
\draw [arrow] (impact) -- (exp);
\draw [arrow] (exp) -- (exppp);
\draw [arrow] (cyberrul) -- (impact);
  
\end{tikzpicture}

%% file: casestudy.tex
\section{Case Study}
\label{sec:casestudy}

In this section we present a case study conducted based on the framework, the case study was intentionally designed to be relatively straightforward, ensuring that it is easily understandable while primarily demonstrating the core functionality of the proposed method. 
In this case study we investigate the use case of remote controllable \ac{HEMS} systems and the impact on the grid when the central control infrastructure of the \ac{HEMS} provider gets compromised.

\subsection{Scenario Description}
\label{subsec:scenario}

The basis for the scenario are combined semi urban medium and low voltage grid from \textit{SimBench}\cite{simbench}.
This benchmark case consists of one 110 kv transformer station and 110 secondary substations connected via a 20 kV network in a ring topology.
Furthermore the low voltage grids have 8982 buses in total connected in radial topologies with 7823 households of which some also have \acp{PV}, \acp{BSS}, heat pumps and \acp{EV} connected to them, additionally commercial loads are part of the model.
The benchmark case also contains production and load time series for the different units, for this paper we analysed the time steps with the highest and lowest sum of transformer loading.
For the augmentation, every bus connected to a household load is defined as household \ac{PCC} and a \textit{errol:HouseHold} gets created, which then owns all units connected to this bus.
If the household contains any controllable loads or production a \ac{HEMS} consisting of a host, a functional actor, and the function block for remote contractility are added, as shown in figure\ref{fig:kg}.
Three different \ac{HEMS} templates for different manufacturers exist, they get added which the probability of 50\%, 30\% and 20\%.
For each of the manufacturers a backend system gets created and the function blocks of the \ac{HEMS} of this manufacturer get connected via an information object flow.
For all controllable units \textit{errol:controlValues} are created referencing the active power of the units, which then get connected to the corresponding information object flow.
Because the objective of this case study is a worst case analysis of the impact of a compromised \ac{HEMS} operator for the grid, the augmentation rules identifies all values which are controllable by a specific backend and either set them to the maximal or minimal possible value.
The underlying assumption for the \ac{BSS} is that in normal operation the state of charge is not fully utilised, because of battery health concerns, and therefore the full power can be used for a short time attack.
We analysed two attacks for each manufacturer one trying to maximise the load in the grid and one trying to minimise it.

\subsection{Result and Discussion}
\label{subsec:result}

\begin{figure}[H]

	\centering
	\includegraphics[width=\linewidth]{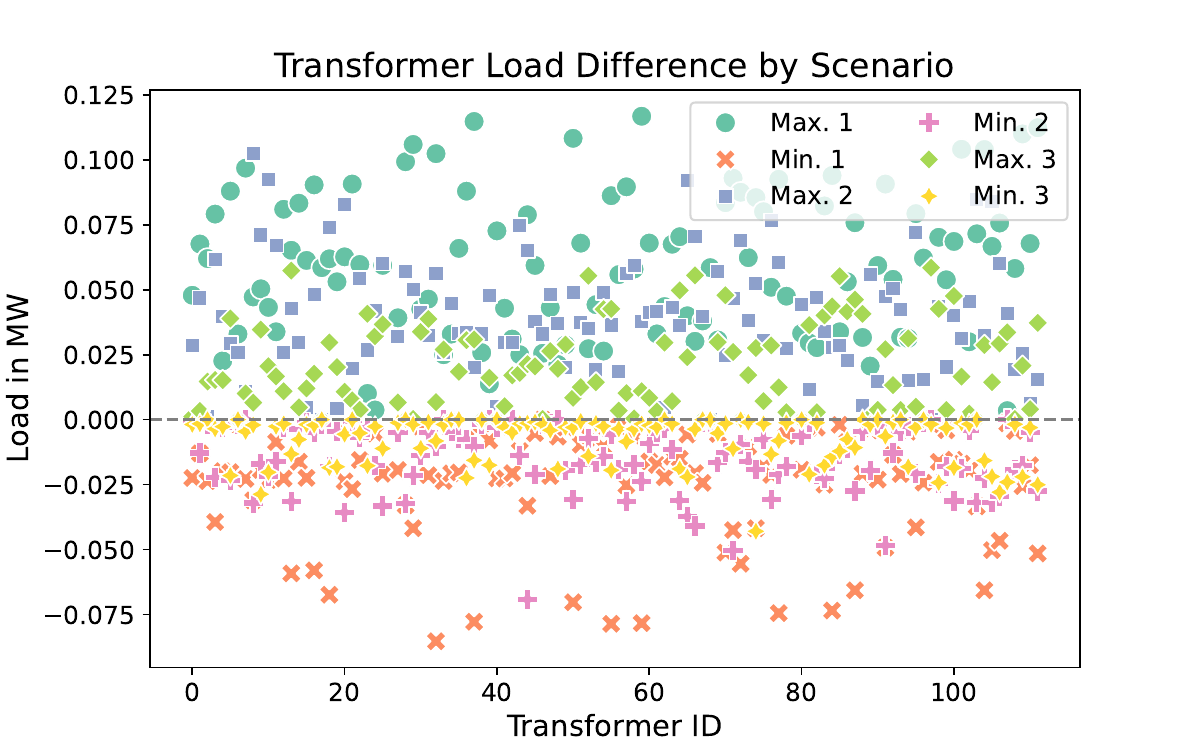}
    \vspace*{-0.3cm} 
	\caption{
		Load difference of transformers in secondary substations for different attack scenarios in MW in the high load case.
	}
	\label{fig:pmw_diff_high}
\end{figure}

Figure \ref{fig:pmw_diff_high} shows the difference in the loading of the transformers of the secondary substations for the high load case.
Every load gets compared to the loading in the scenario with out an attack.
"\textit{Max. 1}", for example, is the difference in load for the scenario that the attacker gains access to manufacture 1 and wants to maximise the load in the grid.
\vspace*{-0.5cm} 
\vspace*{-0.85cm} 
\begin{figure}[H]

	\centering
	\includegraphics[width=\linewidth]{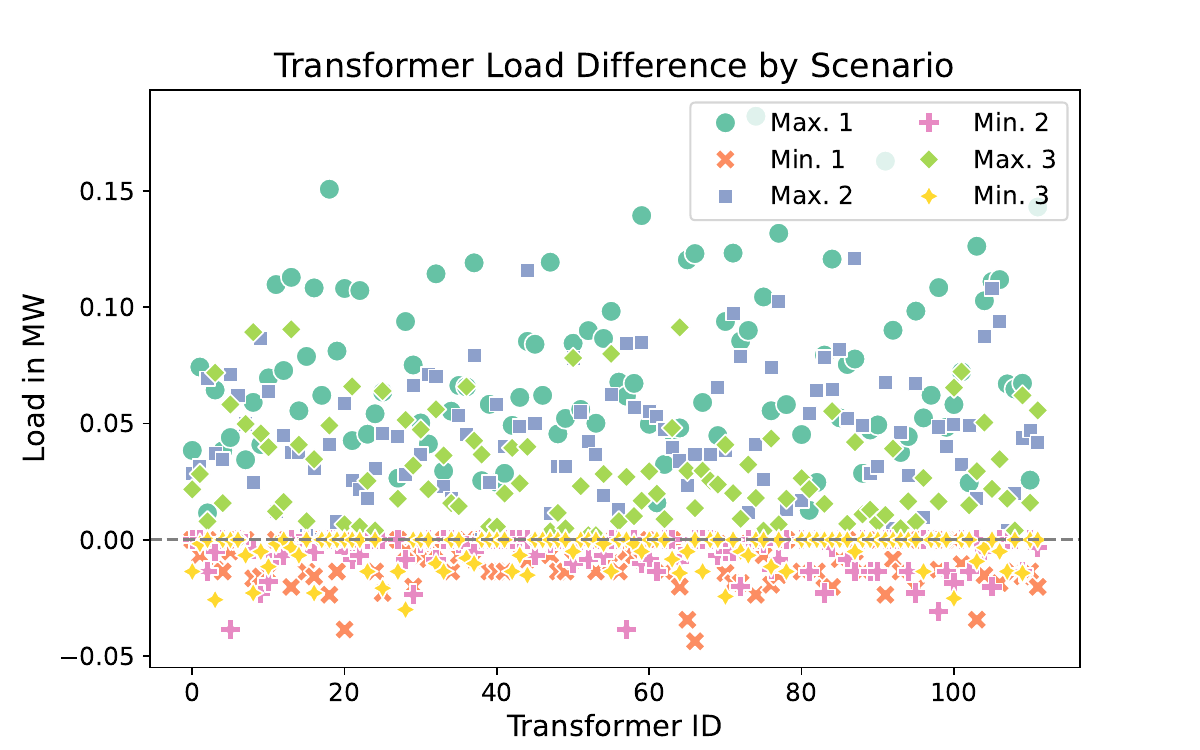}
    \vspace*{-0.3cm} 
	\caption{
		Load difference of transformers in secondary substations for different attack scenarios in MW in the low load case.
	}
	\label{fig:pmw_diff_low}
\end{figure}

\begin{figure}[H]
	\centering
	\includegraphics[width=\linewidth]{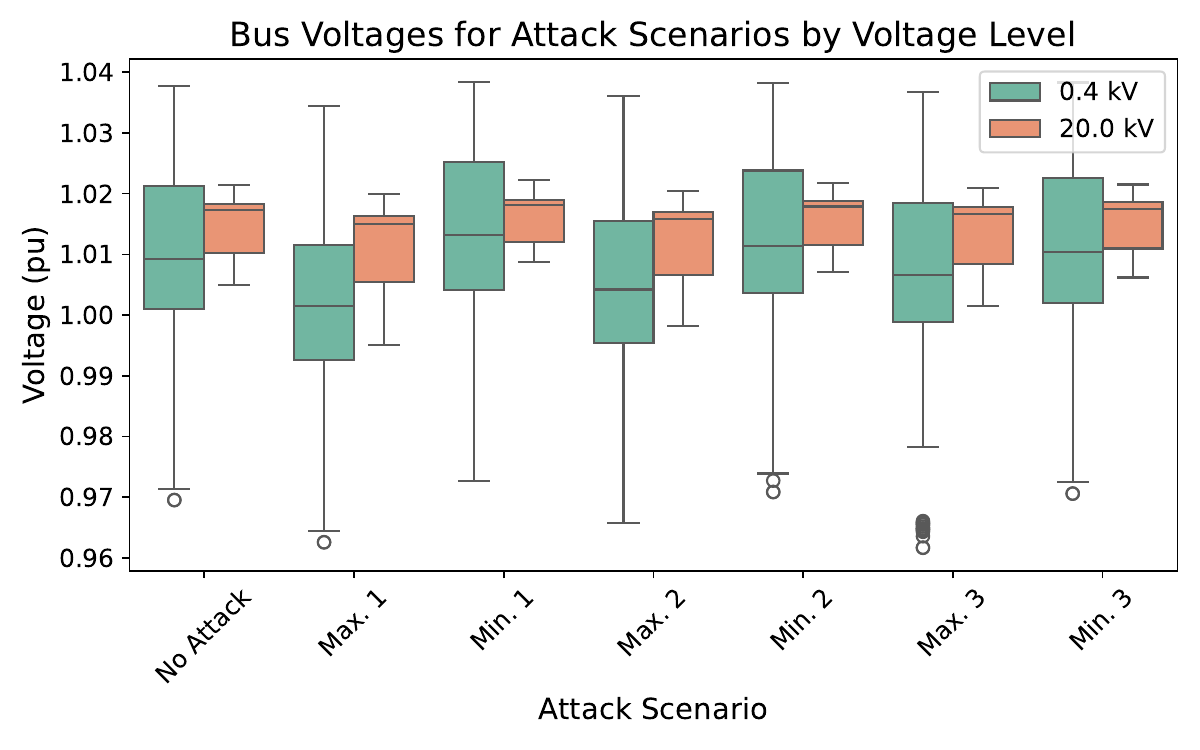}
	\caption{
		Voltages of all busses differentiated by different voltage levels as box plots. For the different attacks scenarios in the high load case.
	}
	\label{fig:box_high}
\end{figure}

Therefore the results for all maximising attacks are above zero and for all minimising attacks are below zero. As expected the attacks on manufacture 1 have the greatest impact in general because it was distributed with the highest probability in the augmentation process.
Nevertheless, for certain substations, this does not hold true because, by coincidence, a larger portion of the load connected to these substations is controllable by a different manufacturer.
Figure \ref{fig:pmw_diff_low} shows the corresponding results for the low load case. When comparing the results it can be seen that the maximising attacks in this case have a slightly greater effect with a load change up to 0.182 MW compared to 0.117 MW.
The delta is mainly caused by turning off the \ac{PV} production.
On the other hand the impact of the minimising attacks is lower because in this case the total power of controllable loads is smaller and \ac{BSS} with the goal of self consumption optimisation are already charging.
In general it can be stated that the impact of the attacks does not lead to problematic grid states, i.e. violation of voltage or line and transformer limits.
This is additionally supported by a specific analysis of the different voltage levels in the scenario.
Figure \ref{fig:box_high} presents all medium and low voltage levels as box plots for the different attack scenarios.
No clear deterioration can be observed, which is consistent with our experience from working with the \textit{SimBench} models, as they exhibit a high robustness to voltage deviations.
However, a study using real network models is necessary to demonstrate the specific impact on the networks.
The presented results nevertheless showcase the functionality of the developed framework and the feasibility of it for these kind of analysis.
\vspace*{-0.4cm} 

%% file: conclusion.tex
\section
{Conclusion and Future Work}

This work presents a comprehensive approach to modelling and analysing the impact of cyber-attacks on interconnected \ac{BTM} infrastructure within a cyber-physical energy system.
By utilising Semantic Web technologies, \ac{SHACL},  we have developed a graph model based on \ac{SGAM} that effectively represents the interdependencies between control infrastructure and the power system.
This model is used in our framework which automatically augments the control infrastructure to the electrical grid and allows the impact analysis of cyber-attacks.
Our case study demonstrates the feasibility of the developed model and framework, showing that the analysed benchmark case is not jeopardised by the examined attack vectors.
Future work will focus on increasing the model's complexity to identify less obvious risks in future energy systems and integrating scenarios that are closer to real-world grid conditions.
Additionally, the framework will be used to generate models of \ac{CPES} as input data for more sophisticated simulation environments, such as co-simulation platforms, providing a foundation for the development of resilience-enhancing measures.